\documentclass[letterpaper, preprint, paper,11pt]{AAS}

\usepackage{bm}
\usepackage{amsmath}
\usepackage{subfigure}
\usepackage{comment}
\usepackage{overcite}
\usepackage{footnpag}			      	
\usepackage{amssymb}
\usepackage{subfigure}   
\usepackage{multicol}
\usepackage{multirow}
\usepackage{booktabs}
\usepackage{pstool}             
\usepackage[colorlinks]{hyperref}
\usepackage{tablefootnote}

\usepackage{amsthm}

\PaperNumber{24-315}

\begin{document}

\title{Parameters Estimation in Source-Sink Space Population Evolutionary Models}

\author{ Erin Ashley\thanks{PhD Candidate, Department of Aerospace Engineering, Iowa State University, IA 50011, USA. email: eashley1@iastate.edu}, 
\ Carla Simon Sanz \thanks{Undergraduate Research Assistant, Department of Aerospace Engineering, Iowa State University, IA 50011, USA. email: carlass@iastate.edu},
\ Simone Servadio\thanks{Assistant Professor, Department of Aerospace Engineering, Iowa State University, IA 50011, USA. email: servadio@iastate.edu}
\ and Giovanni Lavezzi \thanks{Postdoctoral Associate, Department of Aeronautics and Astronautics, Massachusetts Institute of Technology, MA 02139, USA. email: glavezzi@mit.edu}
}

 \maketitle
 
\begin{abstract}
MOCAT-SSEM is a Source-Sink model that predicts the Low Earth Orbit (LEO) space population divided into families using a predefined set of interaction parameters. Thanks to data from the Monte Carlo version of the model (MOCAT-MC), which propagates singularly every object, it is possible to estimate such parameters, assumed as additional stochastic variables. Thus, this paper proposed a new set of parameters so that the new Source-Sink model prediction better fits the computationally expensive and accurate MOCAT-MC simulation. Estimation is performed by extracting stochastic quantities from the space population, which has been analyzed to fit common probability density functions. 
\end{abstract}


\section{Introduction}

Space Domain Awareness (SDA) is an ongoing area of concern as the number of objects in Low-Earth Orbit (LEO) continues to rise [\citen{SDA_threat}, \citen{SDA-SSEM}, \citen{MRIndex}]. Objects in LEO encompass a wide range of sources, functionality, and navigational hazards [\citen{SDA_threat}, \citen{MRIndex}]. Active and derelict satellites, discarded upper rocket stages, and debris from collisions between objects in LEO pose significant navigational hazards for future spacecraft launches and existing orbital infrastructure [\citen{servadio2024risk}]. Even a small number of collisions in LEO can have far-reaching catastrophic effects, injecting hundreds or thousands of new debris into the LEO environment which subsequently need to be tracked and avoided [\citen{NASA_breakup}].
SDA models are important tools in understanding future risk to orbital infrastructure by simulating orbital population evolution and estimating the number and type of objects in LEO in the future [\citen{SDA-SSEM}].

Expansion and refinement of an existing Monte-Carlo SDA model are developed and described in this paper. The MIT Orbital Capacity Analysis Tool (MOCAT) model is a stochastic simulation program that simulates the debris evolution of a full-scale three-dimensional orbital model to estimate the total population of objects by altitude and type classification [\citen{MRIndex}]. The MOCAT model propagates the total number of objects by taking into account the various interactions between each class of object at at each time step. MOCAT is an open-source\footnote{\url{https://github.com/ARCLab-MIT/MOCAT-SSEM}} Source-Sink Evolutionary Model (SSEM) with multiple bins, species, and extended capabilities [\citen{Lifson2023}]. The version of MOCAT-SSEM used in this work defines objects in a three-species classification system; active satellites (S), derelict satellites (D) and debris (N). Though this process is computationally expensive, MOCAT-SSEM is able to distinguish between object species classification and record each object species interaction over the model history, making this an extremely valuable modeling tool. Simulation parameters can also be easily exchanged for different values, allowing the modeling of a range of different orbital scenarios with a single modeling tool.

\begin{figure}
    \centering
    \includegraphics[width=0.95\linewidth]{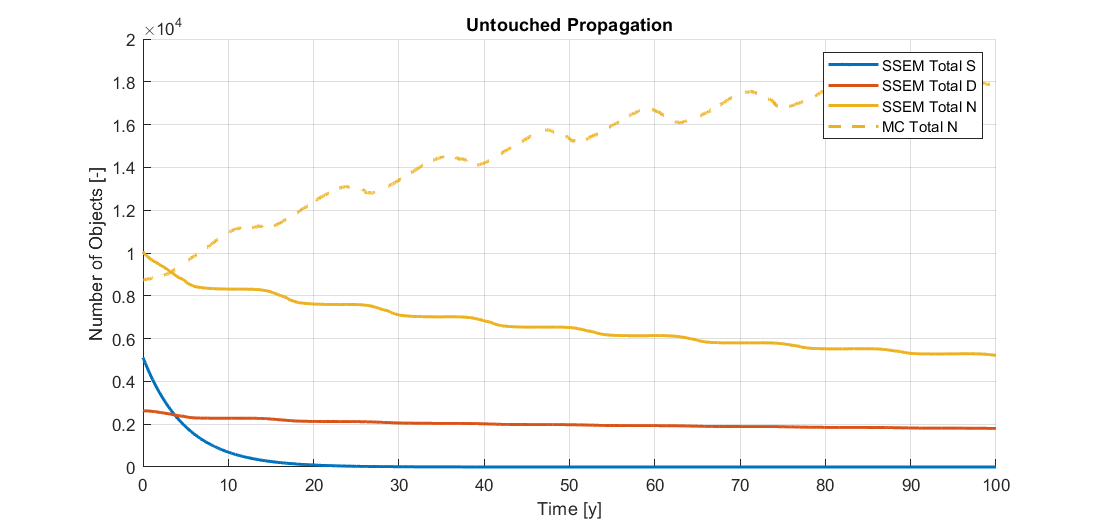}
    \caption{Comparison Between MOCAT-SSEM Object Estimation and MOCAT-MC Object Estimation}
    \label{fig:MOCAT_MC}
\end{figure}

The three-species MOCAT-SSEM model is governed by a set of three equations describing the change in population of each of the three object classes with respect to time [\citen{MRIndex}]. However, one limitation of this model is that these equations assume a static probability of interaction between objects over time, rather than determining the probability of interaction between objects based on the current object population. This paper describes the expansion of the governing equations from three to nine to allow for eventual updating of the interaction probability based on the orbital population at each time step. The key motivation for this study comes from a comparison of a MOCAT generated object evolution over time versus a MOCAT-Monte-Carlo (MOCAT-MC) based object evolution\footnote{\url{https://github.com/ARCLab-MIT/MOCAT-SSEM}} over time Fig.\ref{fig:MOCAT_MC}. For a simulation of 100 years with null new launches using a Jacobian atmospheric model, there is a significant disagreement between the MOCAT-SSEM generated object evolution and a MOCAT-MC based object evolution using a dataset of 4000 TLEs. Due to this significant discrepancy, further refinement of the MOCAT-SSEM model is necessary to ensure a high degree of accuracy.

Another key limitation of the current MOCAT-SSEM model is the lack of measurement updates to the model. For a stochastic system, process noise produces a high amount of uncertainty when an estimate is based on a single measurement [\citen{EstimationTracking}]. Kalman filtering is a commonly used technique to update the predicted state of a system by comparing the current estimate to a measurement of the system at each time step [\citen{K_LinearFiltering}]. The application of Kalman filtering in guidance, navigation, and control applications is well documented [\citen{Kalman_History}, \citen{SR_UKF}]. Unscented Kalman filtering is a robust method of evaluating uncertainty when dealing with significantly non-linear systems such as those commonly found in astrodynamics [\citen{Uncertaintiy_Filter}, \citen{ODetFundam}, \citen{EstimationConditional}]. This research describes the implementation of a Unscented Kalman filter (UKF) update step to the MOCAT-SSEM model, improving the accuracy of the simulation results. Future work will use an Extended Kalman Filter to apply data smoothing to the Unscented Filter results.
Another important contribution of this paper is the development of a performance index for different distribution fitting parameters. The performance index is determined by applying root-mean-square error (RMSE) analysis on MOCAT-MC simulation data when fitted with a series of distribution parameters.

\section{Methodology}

\subsection{Overview of the MOCAT Model and Simulation Parameters}
The MOCAT-MC model is a Monte-Carlo based approach to simulating the LEO environment. The initial orbital population data set\footnote{\url{https://github.com/ARCLab-MIT/MOCAT-SSEM}} draws from several sources to define the number type, and characteristics of each object. Identification of number of objects currently in orbit is sourced from Space-Track\footnote{\url{https://www.space-track.org}} data. Object parameters, such as object-class, mass, diameter, and launch date, are sourced from the European Space Agency developed database, DISCOSweb\footnote{\url{https://discosweb.esoc.esa.int}}. Whether an object can perform stationkeeping operations is determined by its active or inactive status. Object status information is sourced from CelesTrak\footnote{\url{https://celestrak.org}}. These sets of object parameters are recorded in Two Line Elements sets (TLEs). TLE seed numbers are used to index simulation results when analyzing data. The three governing equations for each shell, $i$, of the current MOCAT model describe the change in orbital population per species with respect to time [\citen{SSEM_Orbit_Capacity}]. The drag coefficient for active and inactive satellites is assumed to be 2.2.

\begin{equation}
\dot{S}^{(i)} = \lambda^{(i)} - \frac{S^{(i)}}{TOF} - \alpha_a \phi_{SS} {S^{(i)}}^{2} - (\delta + \alpha) \phi_{SD} S^{(i)} D^{(i)} - (\delta + \alpha) \phi_{SN} S^{(i)} N^{(i)} \label{Eq. 1}
\end{equation}

\begin{equation}
\dot{D}^{(i)} = \frac{(1-PMD) S^{(i)}}{TOF} - \delta S^{(i)}D^{(i)} + \delta S^{(i)} N^{(i)} - \phi_{DD} {D^{(i)}}^{2} - \phi_{DN} D^{(i)} N^{(i)} + {\dot{F}_{d,D}}^{(i + 1)} - {\dot{F}_{d,D}^{(i)}} \label{Eq. 2}
\end{equation}

\begin{equation}
\begin{split}
\dot{N}^{(i)} &= n_{f,SS} \alpha_a \phi_{SS} {S^{(i)}}^{2} + n_{f,SD} \alpha \phi_{SD} S^{(i)} D^{(i)} + n_{f,SN} \alpha \phi_{SN} S^{(i)} N^{(i)} \\
&+ n_{f,DD} \phi_{DD} {D^{(i)}}^{2} + n_{f,DN} \phi_{DN} D^{(i)} N^{(i)} + n_{f,NN} \phi_{NN} {N^{(i)}}^{2} \\
&+ {\dot{F}_{d,N}}^{(i + 1)} - {\dot{F}_{d,N}}^{(i)} \label{Eq. 3}
\end{split}
\end{equation}

Probability of interaction between objects is approximated by the kinetic theory of gases. Each term $\phi_{XY}$ in the three governing equations describes the likelihood of interaction between objects of species $X$ and $Y$, and is determined by [\citen{SSEM_Orbit_Capacity}]:
\begin{equation}
\phi_{XY}^{(i)} = \frac{\pi * {v_{rel}}^2 * (r_X - r_Y)}{V^{(i)}} \label{Eq. 4}
\end{equation}
where $v_{rel}$ is the relative velocity between objects, $r_X$ and $r_Y$ are the mean radii of objects in species $X$ and $Y$, respectively, and where $V^{(i)}$ is the volume of orbital shell $i$. All $\phi_{XY}$ parameters used in the MOCAT-SSEM and MOCAT-MC models are static with respect to orbital population.
For objects with active control elements, the parameters $\alpha_{a}$ and $\alpha$ refer to the probability that a collision will occur between two objects with active control, and one object with active control and one without, respectively. By default, these parameters are established at 0.01 and 0.2, respectively. Ratio of disabling to lethal debris is given by $\delta$ and is set to ten in MOCAT-SSEM.

An atmospheric drag model is used to determine the rate at which derelict and debris objects of class $X$ decay $\dot{F}_{d,X}$ from the orbital shell above into the orbital shell below. The current MOCAT model allows for interchanging of atmospheric models. The number of new spacecraft launches, $\lambda$, can be defined at either a constant rate or a null rate. Simulation results analyzed in this paper were obtained by using a JB2008 atmospheric model accounting for peak cycle solar activity and null rate new launches. Simulation results with a Gaussian distribution of new launches into all shells and a JB2008 atmospheric model were used in the UKF in addition to results using null rate new launches for further analysis. Time of flight, $TOF$, is averaged at five years for active objects in LEO. The probability of successful post-mission disposal, $PMD$, is assumed to have a success rate of 0.95. The number of new fragments generated by collisions between objects of classes $X$ and $Y$ are based on the NASA standard breakup model of collisions, appearing in Eq.\eqref{Eq. 3} as terms $n_{f,XY}$ [\citen{NASA_breakup}].

A sample of 4,000 MOCAT-MC simulation results was used for the development of distribution fitting parameters described in this work. These simulations provide the data set used as measurement mean and covariance for the Kalman filter update. In this work, the number of orbital shells is equal to thirty-six, beginning at an altitude of 200 km and ending at an altitude of 2000 km. The height of each orbital shell is assumed to be 50 km.

\subsection{Data Extraction and Visualization}
Data extraction and visualization consisted of extracting the Monte-Carlo population data for 4000 simulations and calculating the mean, covariance, and cross-covariance between the number of objects over the number of simulations at each time step. Due to the large volume of data being analyzed, the mean and covariance calculations were run for only a single shell at a time for computational efficiency. This process was repeated for each orbital shell $i$.

\begin{align}
G^{i} = \begin{bmatrix}
        {\hat{S}}^{i} \\
        {\hat{D}}^{i} \\
        {\hat{N}}^{i} \\
        {cov(SS)}^{i} \\
        \vdots \\
        {cov(NN)}^{i} \\
        \end{bmatrix}
\end{align}

\begin{figure}[!ht]
    \centering
    \includegraphics[width=0.99\linewidth]{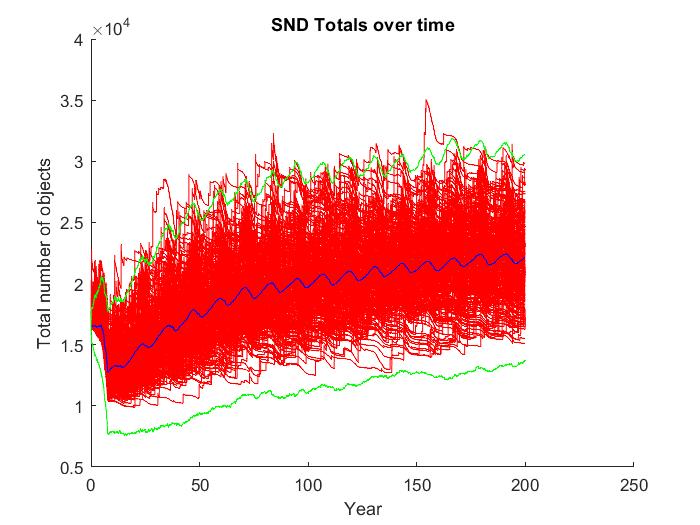}
    \caption{Evolution of S, N, D, For All Shells With Total Mean (Blue), 3$\sigma$ Deviation (Green), and TLE Simulations (Red)}
    \label{fig:SND_Tot}
\end{figure}

The results of the data extraction process for all thirty-six orbital shell are shown in Fig. \ref{fig:SND_Tot}. Each red line represents a different MOCAT-MC simulation. For visualization purposes, only 250 of the 4000 TLEs are plotted, as the overall mean and covariance behavior slightly change.

While concentrated around the mean, a significant number of MOCAT simulations resulted in many catastrophic collisions, injecting much greater than normal amounts of debris into the LEO environment, and causing the number of objects to frequently exceed the positive $3\sigma$ bound. These catastrophic collisions coincide roughly with the peak solar activity years of the J2008 atmospheric model.
To best visualize the behavior of the object distribution, the probability density functions (PDFs) at several time steps where object distribution was heavily skewed were extracted and plotted with several different distribution fitting parameters. This data can be seen in the discussion of the fitting parameter performance index.

\subsection{Performance Index of Fitting Parameters}
\begin{figure}[!ht]
    \centering
    \includegraphics[width = 0.49\linewidth]{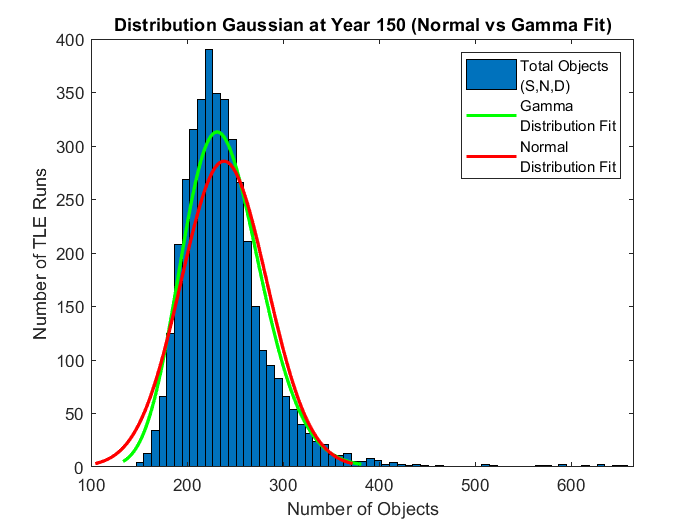}
    \includegraphics[width = 0.49\linewidth]{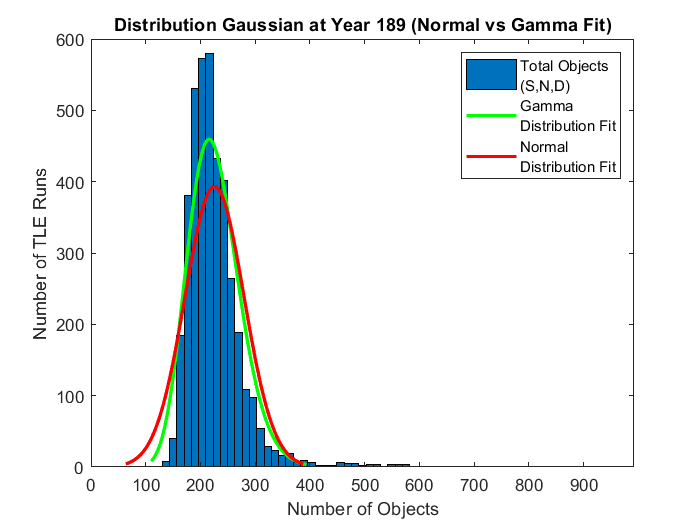}
    \caption{Comparison of Normal and Gamma Fitting for S, N, D, in Shell 8 at Year 150 (Left) and Year 189 (Right)}
    \label{fig:hist150_189}
\end{figure}

Due to the non-normal distribution of object population across MOCAT simulations, it is important to determine the most accurate distribution fitting method. In this work we contribute an analytically determined performance index of different distribution fitting methods. An example is shown in Fig. \ref{fig:hist150_189} for the distribution of total objects of shell 8 at year 150 and year 189,  fitted with a normal and a gamma distribution.

To determine the performance index of each fitting method, a Root-Mean-Square-Error (RMSE) approach is utilized. After normalizing the data such that both the total population histogram and the area under the distribution fitting curve are both equal to one, the height difference between the center of each bin, $X_i$, and the corresponding point along the fitted curve, $\hat{X}_i$ is taken and used to determine the RMSE value for a given distribution fit.
\begin{equation}
RMSE = \sqrt{\frac{\sum_{i = 1}^{N} (X_i - \hat{X}_i)^2}{N}} \label{Eq. 6}
\end{equation}
where $N$ is the number of bins in the histogram plot. The closer the RMSE value for each fit is to zero, the more accurate distribution approximates the actual population spread. In this work, we contribute a performance index for Gaussian, Gamma, and Rician distribution fitting for the total number of objects at each time step in each orbital shell.

\subsection{Expanding the MOCAT System Of Equations}
As mentioned above, the current MOCAT model is governed by a system of three equations. Each of these equations describes the change in species population at each time step approximated from the kinetic theory of gases. However, the accuracy of these predictions is limited by the static parameters $\phi_{XY}$. For more accurate results, the probability of collision between objects needs to be determined at each time step based on the updated number of objects in each shell. We begin by augmenting the system of governing equations from three to nine.

\begin{align}
\dot{X} = \begin{bmatrix}
        \dot{S}_{1-36} \\
        \dot{D}_{1-36} \\
        \dot{N}_{1-36} \\
        \end{bmatrix} \rightarrow
        \begin{bmatrix}
        \dot{S}_{1-36} \\
        \dot{D}_{1-36} \\
        \dot{N}_{1-36} \\
        \dot{\phi_{SS}}_{1-36} \\
        \vdots \\
        \dot{\phi_{NN}}_{1-36} \\
        \end{bmatrix}
\end{align}

\begin{figure}
    \centering
    \includegraphics[width = 0.49\linewidth]{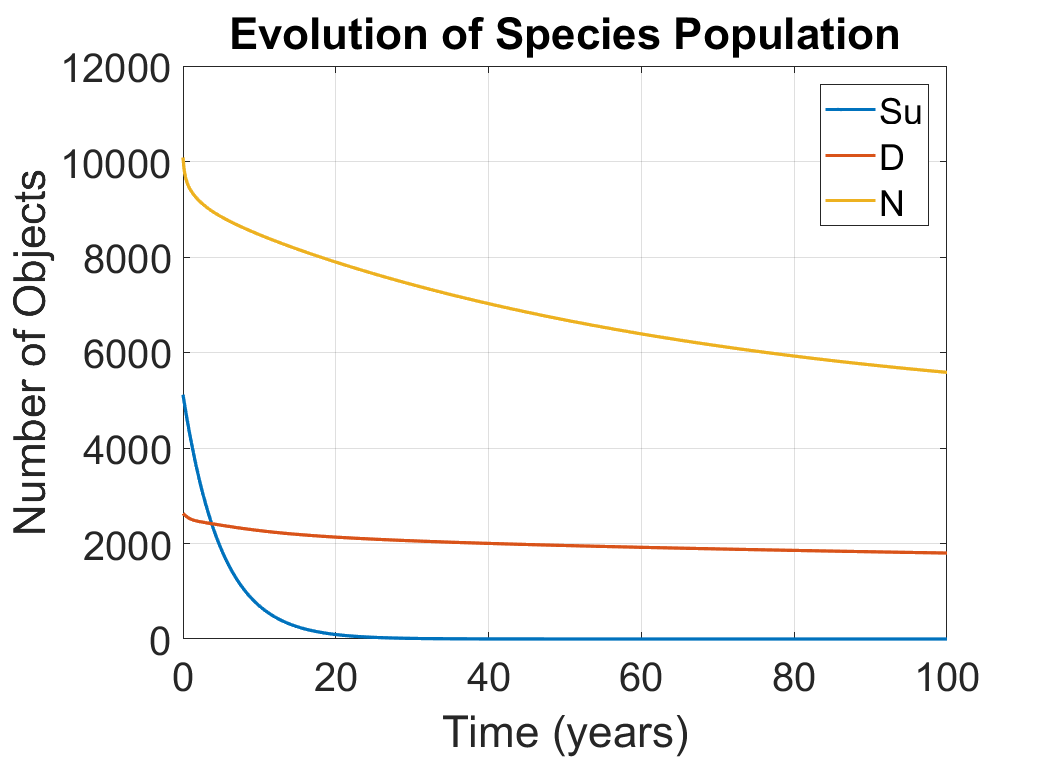}
    \includegraphics[width = 0.49\linewidth]{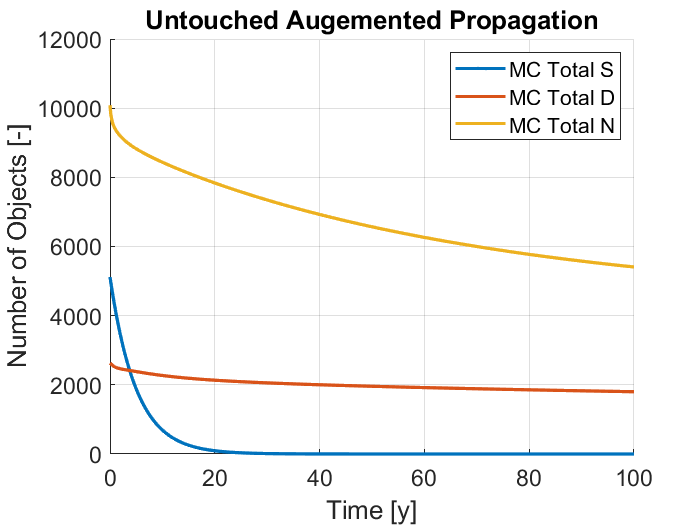}
    \caption{Time Propagation of Unmodified System of ODEs (Left) and Augmented System of ODEs (Right)}
    \label{fig:prop}
\end{figure}

Applying these changes to all 36 orbital shells results in the state vector used in the augmented system of equations expanding from 108 to 324. Initial testing of this augmented system dynamics vector was performed by setting each of the $\dot{\phi}_{XY}$ equations equal to zero ($\phi_{XY}$ is static) and comparing propagation of the augmented state with the propagation of the non-augmented state, Fig. \ref{fig:prop}. These results show that the augmented state vector propagation is able to replicate the results obtained by the three-species MOCAT-SSEM when input conditions are the same. From this, we can ensure that the propagation functions for S, D, and N are behaving correctly.

\subsection{Developing the Kalman Filter Update}
Due to the significant non-linearity of the system dynamics, filtering is performed by application of a UKF [\citen{Uncertaintiy_Filter}, \citen{ODetFundam}]. The UKF method provides significantly greater accuracy for systems that cannot be linearly approximated [\citen{NonlinearEstimation}, \citen{NonlinearTransformation}].
Initialization of the UKF defines $x_0$ as the initial state of the MOCAT-MC mean, and defines $P_0$ as a diagonal matrix with values equal to ten percent of the initial state estimate.
The state transition function and measurement function used in the UKF are as follows:

\begin{equation}
    f(\chi_k) = \begin{bmatrix}
        \dot{S}_{1-36}(\chi_k) \\
        \dot{D}_{1-36}(\chi_k) \\
        \vdots \\
        \dot{\phi_{NN}}_{1-36}(\chi_k) \\
    \end{bmatrix}
    \label{stfunc}
\end{equation}

\begin{equation}
    h(\chi_k) = f(\chi_k)
    \label{measfunc}
\end{equation}
and the UKF follows the algorithm below.

Step 1. Calculation of Sigma Points $\chi^{(i)}$
\begin{equation}
    \chi_k = \bigg[ \hat{x}_{k|k}  \quad \hat{x}_{k|k}+\sqrt{(n+\lambda)P_{k|k}}  \quad \hat{x}_{k|k}-\sqrt{(n+\lambda)P_{k|k}} \bigg] \label{Eq. 7}
\end{equation}
where $\hat{x}_{k|k}$ is the estimate mean at time $k$ updated with the measurement at time $k$, $P_{k|k}$ is the estimate covariance at time $k$ updated with the measurement at time $k$.
The total number of sigma points is $n = 2(n_{states}) + 1$ and $\lambda = a^2 (n_{states} + \kappa) - n_{states}$ is a scaling parameter. The spread parameter $a$ and secondary scaling parameter $\kappa$ are equal to 0.25 and 0, respectively.

Sigma point weights of the estimate and covariance are calculated by the equations
\begin{equation}
    W_{m(0)} = \frac{\lambda}{(n_{states}+\lambda)}
    \label{Wm0}
\end{equation}
\begin{equation}
    W_{c(0)} = \frac{\lambda}{(n_{states}+\lambda)} + (1-a^2+\beta)
    \label{Wc0}
\end{equation}
\begin{equation}
    W_{m(i)} = W_{c(i)} = \frac{1}{2(n_{states}+\lambda))}
    \label{weights}
\end{equation}
where $\beta$ is a secondary parameter used to incorporate prior knowledge of the distribution. $\beta = 2$ is optimal for Gaussian distributions.

Step 2. Propagation of state estimate mean and covariance
\begin{equation}
    \chi^{(i)}_{k+1} = f(\chi^{(i)}_{k}) \rightarrow i = 0,...,2n \label{Eq. 8}
\end{equation}
\begin{equation}
    \hat{x}_{k+1|k} = \sum^{2n}_{i=0} W^{(i)}_{m} \chi^{(i)}_{k+1} \label{Eq. 9}
\end{equation}
\begin{equation}
    P_{k+1|k} = \sum^{2n}_{i=0} W^{(i)}_{c} (\chi^{(i)}_{k+1}-\hat{x}_{k+1|k}) (\chi^{(i)}_{k+1}-\hat{x}_{k+1|k})^T + Q \label{Eq. 10}
\end{equation}
with additive process noise equal to five percent of the initial state estimate being represented by $Q$ and the propagated estimate and covariance represented as $\hat{x}_{k+1|k}$ and $P_{k+1|k}$.

Step 3. Update
\begin{equation}
    y^{(i)} = h(\chi^{(i)}_{k+1}) \rightarrow i = 0,...,2n \label{Eq. 11}
\end{equation}
\begin{equation}
    \hat{y} = \sum^{2n}_{i=0} W^{(i)}_{m} y^{(i)} \label{Eq. 12}
\end{equation}
\begin{equation}
     P_{yy} = \sum^{2n}_{i=0} W^{(i)}_{c} (y^{(i)}-\hat{y}) (y^{(i)}-\hat{y})^T + R \label{Eq. 13}
\end{equation}
\begin{equation}
    P_{xy} = \sum^{2n}_{i=0} W^{(i)}_{c} (\chi^{(i)}_{k+1}-\hat{x}_{k+1|k}) (y^{(i)}-\hat{y})^T \label{Eq. 14}
\end{equation}
\begin{equation}
    K = P_{xy} P^{-1}_{yy} \label{Eq. 15}
\end{equation}
\begin{equation}
    \hat{x}_{k+1|k+1} = \hat{x}_{k+1|k} + K(y-\hat{y}) \label{Eq. 16}
\end{equation}
\begin{equation}
    P_{k+1|k+1} = P_{k+1|k} - K P_{yy} K^T \label{Eq. 17}
\end{equation}
where $y^{(i)}$ are the sigma points associated with the measurement. In this work the measurement is the extraction of the species S, D, and N from the state. Thus the measurement function $h$ is linear as given in Eq. \eqref{measfunc}. Where $\hat{y}$ is the measurement at time $k+1$, and $R$ is the measurement covariance obtained from the data extraction of the MOCAT-MC results. The terms $P_{xy}$ and $P_{yy}$ refer to the covariance between the measurement and itself and the covariance between the estimate and the measurement, respectively. Parameter $K$ is the Kalman Gain, $y$ is the measurement at time $k+1$, $\hat{x}_{k+1|k+1}$ refers to the updated estimate at time $k+1$ using the measurement at time $k+1$, and $P_{k+1|k+1}$ refers to the updated covariance of the estimate at time $k+1$ using the measurement at time $k+1$.

Measurement estimates and covariances used in the Kalman filter are referenced from the data extraction of each 4000 TLE subset. Untouched measurements used in the Kalman filter are referenced directly from the untouched system propagation Eq. \eqref{Eq. 10}. The measurement vector is organized by shell. The measurement covariance matrix is similarly organized blockwise by shell.

\begin{align}
        \hat{y} = \begin{bmatrix}
        \hat{S}_1 \\
        \hat{D}_1 \\
        \hat{N}_1 \\
        \vdots \\
        \hat{S}_{36} \\
        \hat{D}_{36} \\
        \hat{N}_{36} \\
        \end{bmatrix}
        R_{ij} = \begin{bmatrix}
        P_{SS_{ij}} & P_{SD_{ij}} & P_{SN_{ij}} \\
        P_{SD_{ij}} & P_{DD_{ij}} & P_{DN_{ij}} \\
        P_{SN_{ij}} & P_{DN_{ij}} & P_{NN_{ij}} \\
        \end{bmatrix}
\end{align}

Where $i$ and $j$ are the shells for which the covariances are calculated. When $i \neq j$, the covariances are between shells as well as between object types.

\section{Preliminary Results}
\subsection{Performance Index Results}
\begin{figure}[!ht]
    \centering
    \includegraphics[width = 0.46\linewidth]{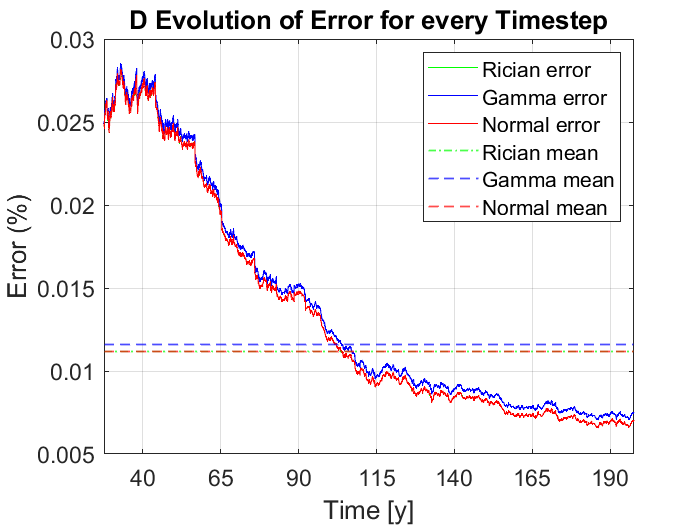}
    \includegraphics[width = 0.52\linewidth]{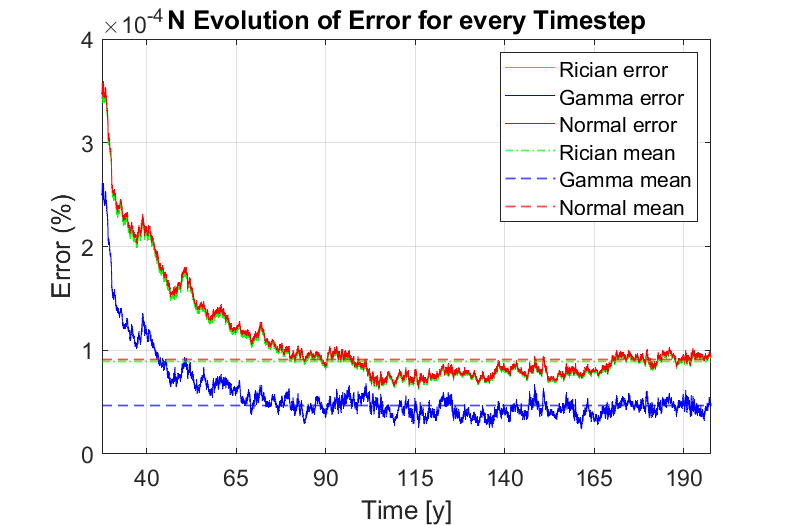}
    \caption{Error Comparison for Gaussian, Gamma, and Rician RMSE Fits in Shell 16 for Derelict Objects (Left) and Debris (Right)}
    \label{fig:D16_N16}
\end{figure}

RMSE values for Gaussian, Gamma, and Rician histogram fitting curves were analyzed at each time step for individual shells. The process was repeated for each shell and the results plotted over time. An example for RMSE fits in shell 16 is provided in Fig. \ref{fig:D16_N16}. Shell 16 is selected for visualization due to the high number of total objects populating this shell consistently throughout the simulation. General performance index formulae are given.

\begin{equation}
    _{Gaussian}\rho_{i,k} = \sqrt{\frac{\sum_{i = 1}^{N} (X_{i,k}^{Gaussian} - \hat{X}_{i,k})^2}{N}}
    \label{Eq. 24}
\end{equation}
\begin{equation}
    _{Gamma}\rho_{i,k} = \sqrt{\frac{\sum_{i = 1}^{N} (X_{i,k}^{Gamma} - \hat{X}_{i,k})^2}{N}}
    \label{Eq. 25}
\end{equation}
\begin{equation}
    _{Rician}\rho_{i,k} = \sqrt{\frac{\sum_{i = 1}^{N} (X_{i,k}^{Rician} - \hat{X}_{i,k})^2}{N}}
    \label{Eq. 26}
\end{equation}
where subscript $i$ indicates the specific simulation time $i$ while subscript $k$ indicates the orbital shell. To provide an overall parameter that represents the level of accuracy of a distribution fitting for the entire simulation, we take the expected value of the performance index over time for each shell $n$:
\begin{equation}
    \bar{\rho}_n = \mathbb{E}_{t}[\rho_{i,n}] =  \frac{1}{T}\Sigma_i^T \rho_{i,n}
    \label{Eq. 27}
\end{equation}
where $\mathbb{E}_{t}$ is the expected value operator and $T$ is the total number of time steps considered. 

\begin{figure}[!ht]
    \centering
    \includegraphics[width=0.49\linewidth]{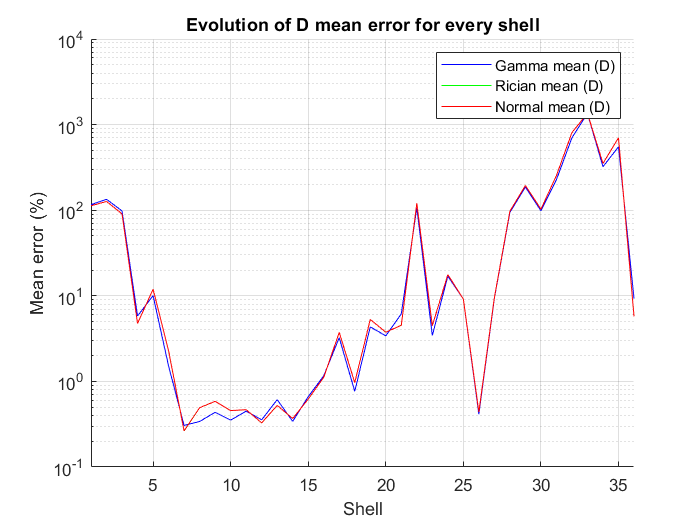}
    \includegraphics[width=0.49\linewidth]{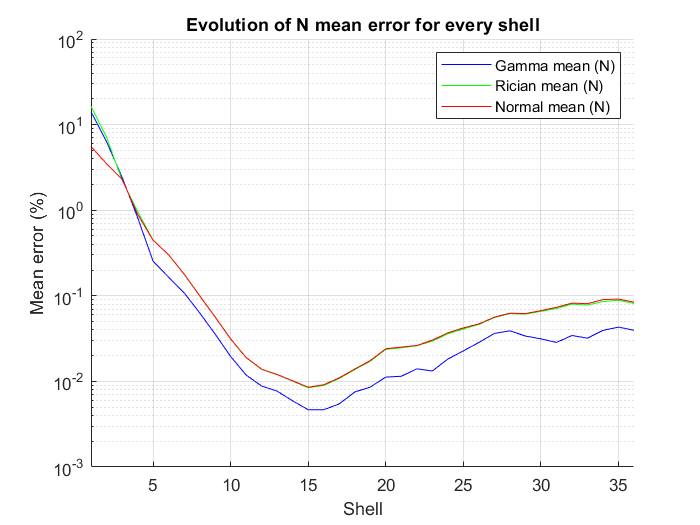}
    \caption{Mean Error Comparison for Gaussian, Gamma, and Ricean RMSE Fits Across All Shells for Derelict Objects (Left) and Debris (Right)}
    \label{fig:Dall-Nall}
\end{figure}

\begin{figure}[!ht]
    \centering
    \includegraphics[width=0.5\linewidth]{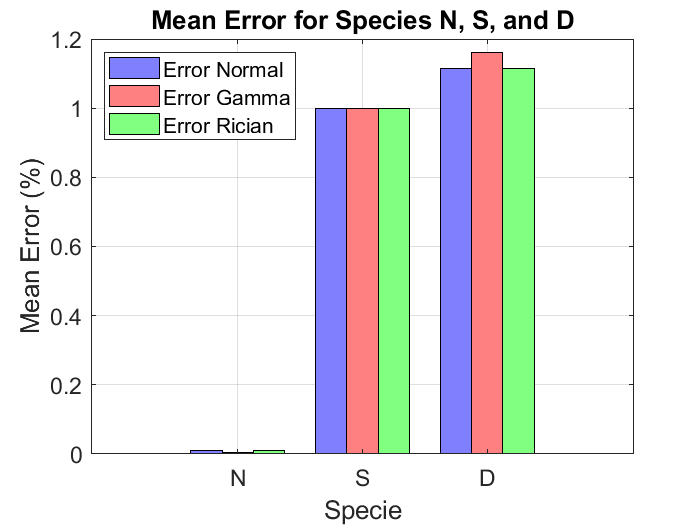}
    \caption{Mean RMSE Error Visualized by Object Species}
    \label{fig:rmse_by_species}
\end{figure}

As usual, a lower RMSE value indicates a more accurate fit of the distribution curve to the data. After establishing RMSE fitting parameters over time for each shell, the steady-state RMSE means over time for each shell were combined for visualization of RMSE fitting parameters across all shells Fig. \ref{fig:Dall-Nall}. Further analysis of RMSE error by object type was also conducted Fig. \ref{fig:rmse_by_species}.

The figure shows that for a single highly populated shell, the Gamma distribution better fits the data set of N when compared with Gaussian and Rician distributions. However, Gaussian distribution fitting more accurately conforms to the data set of D for the same highly populated shell, though the difference between the Gaussian, Gamma, and Rician distributions was much smaller than for N.

When comparing RMSE of the distribution results across all shells, we observed higher error in the fitting parameters for lower altitude shells. This may be explained by the much smaller populations in these orbits. While the Gaussian distribution initially proved more accurate for lower altitude shells, the Gamma distribution proved to be more accurate from shell three and above. This again may be attributed to the smaller populations in lower orbits making it difficult to obtain a proper distribution curve.

Rician distributions experienced similar difficulty in conforming to the data of smaller population shells. This is especially clear for D in the lowest and highest shells, where there may be few or no objects. Population data for D is also significantly smaller than the population data for N, and this smaller population likely exacerbates the issue by making it more difficult to obtain a proper distribution curve.

The significant increase in fitting parameter accuracy for N when compared to S and D is expected due to the dominance of N in the overall orbital population. Due to this discrepancy, we conclude that the performance index for object type N is a good indicator of fitting parameter accuracy of the entire system, and that a Gamma fit better represents the distribution of objects in LEO. Since explosions and collisions behave more like gas particles where one tail of the distribution is more prominent, this result is expected given the distribution as seen in Fig. \ref{fig:SND_Tot} and Fig. \ref{fig:hist150_189}. We also hypothesize that the sample size for D is not significant enough to form an accurate distribution, leading to much higher error percentage when compared to the sample of N as observed in Figs. \ref{fig:D16_N16}, \ref{fig:Dall-Nall}. Initial setup of the UKF filter was performed using Gaussian distribution sigma point calculation to establish a baseline. Future efforts will use Gamma distribution sigma point calculation for sigma point calculation and compare estimation results for further analysis.

\subsection{UKF Estimation Results}
Due to null shells, in order to avoid singularities in the update matrix, we assume a fixed minimum uncertainty for all object types in each shell. Minimum measurement covariance is assumed to be one object, and minimum estimate covariance is assumed to be 0.1, or ten percent of one object. The updated covariance is checked to ensure no numerical problems arise due to empty shells.

\begin{figure}[!ht]
    \centering
    \includegraphics[width=0.99\linewidth]{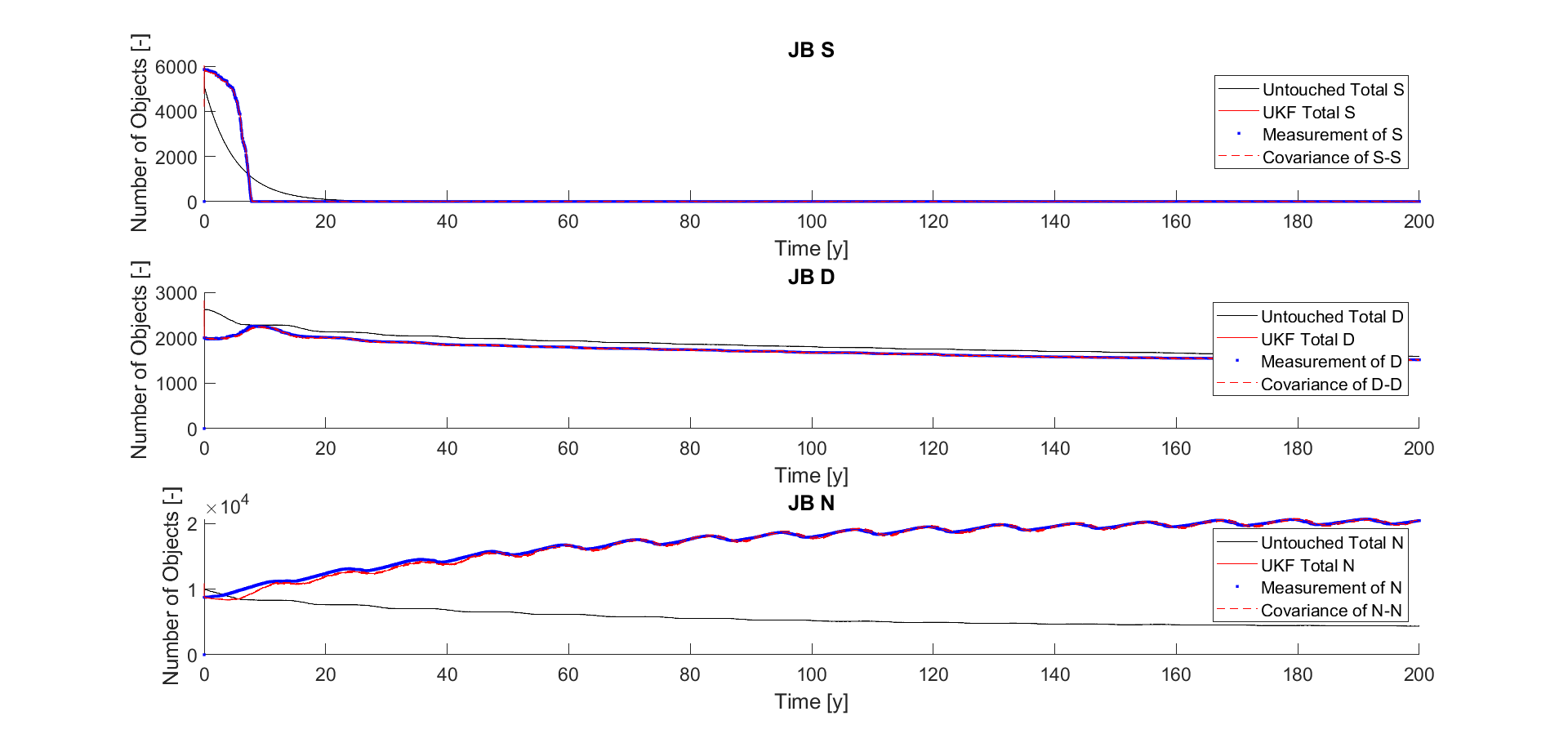}
    \caption{Object Population Evolution Using High-Fidelity JB2008 Atmospheric Case With Null New Launches}
    \label{fig:JB_SDN}
\end{figure}

\begin{figure}[!ht]
    \centering
    \includegraphics[width=0.99\linewidth]{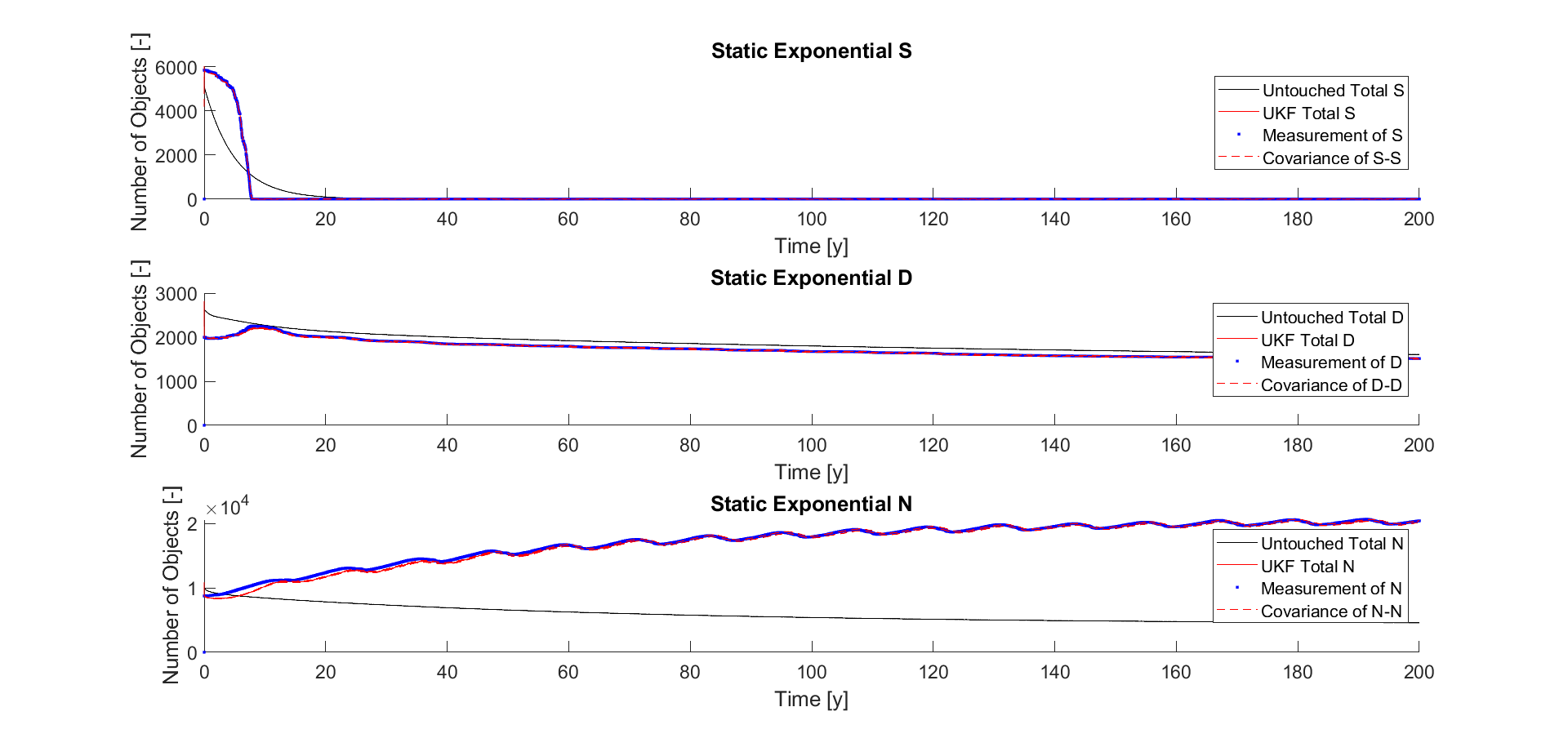}
    \caption{Object Population Evolution Using Low-Fidelity Static Exponential Atmospheric Case With Null New Launches}
    \label{fig:SE_SDN}
\end{figure}


For comparison, a UKF filter with no modifications was run alongside a UKF filter which only considers the real components of the estimate and covariance. Both simulations were run for a total of forty years, with no significant difference in the estimate or covariance. We hypothesize that complex components are produced during each subsequent update due to the difference in magnitude between the object estimates and the phi estimates being incorporated into the same system state, making the augmented state vector ill-conditioned for use in a UKF filter. Results presented in this paper use only the real components of the estimate, as seen in Fig. \ref{fig:JB_SDN} and Fig. \ref{fig:SE_SDN}.

One high-fidelity UKF filter and one low-fidelity UKF filter were simulated and the results compared. The low-fidelity case uses a static exponential atmospheric model for propagation of the state estimate and a set of measurements from a Jacobian atmospheric model. The high-fidelity case used a Jacobian atmospheric model both for propagation of the state estimate and measurement vectors. We observed a negligible difference in the total number of each object type between the high-fidelity and low-fidelity cases, and the measurement values stayed largely within the $3\sigma$ bounds of the estimate covariance for each object type. However, noticeable differences in the estimate evolution occurred between the high-fidelity and low-fidelity cases.

\begin{figure}[!ht]
    \centering
    \includegraphics[width=0.99\linewidth]{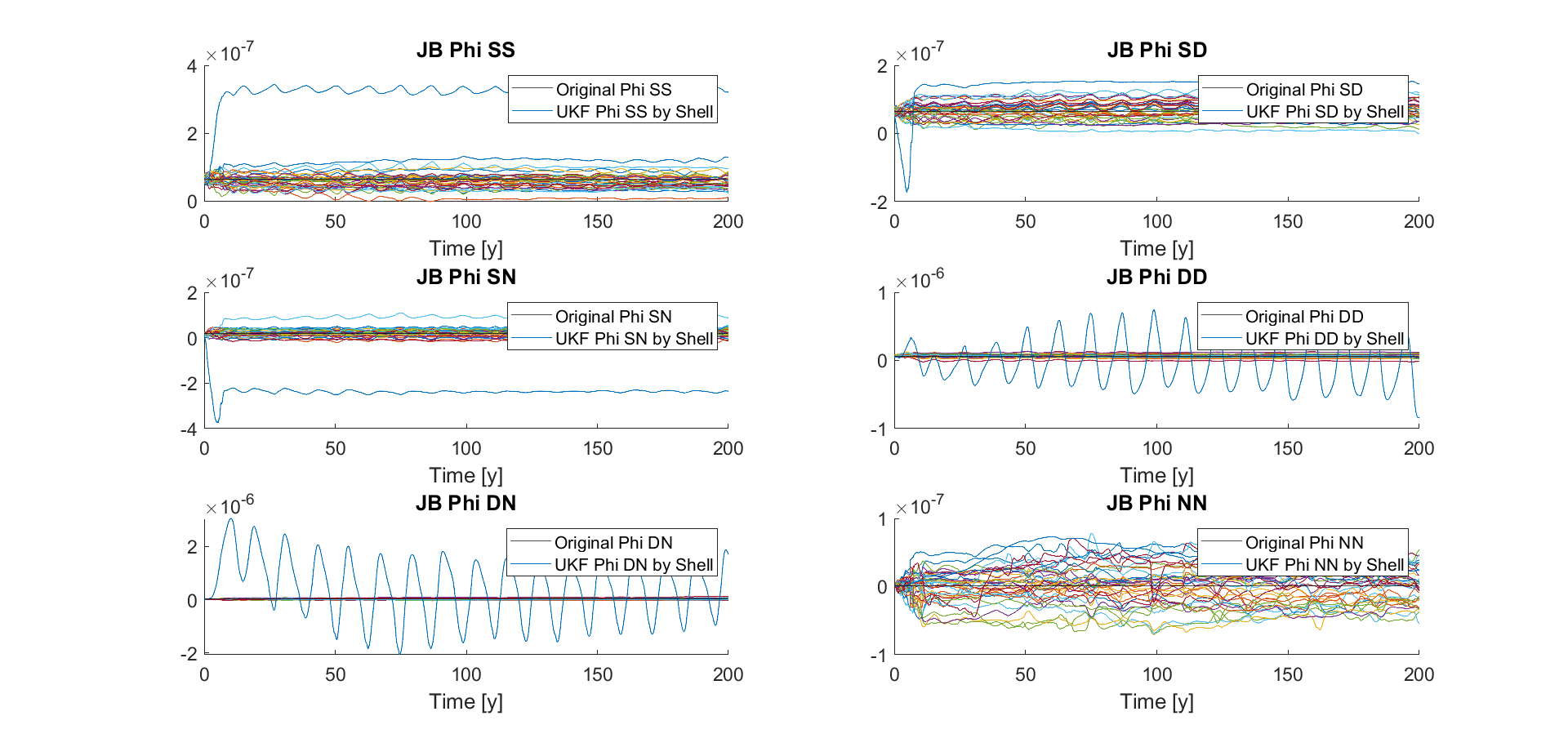}
    \caption{Phi Value Evolution Over Time for JB2008 Atmospheric Model}
    \label{fig:JB_phi_all}
\end{figure}

\begin{figure}[!ht]
    \centering
    \includegraphics[width=0.99\linewidth]{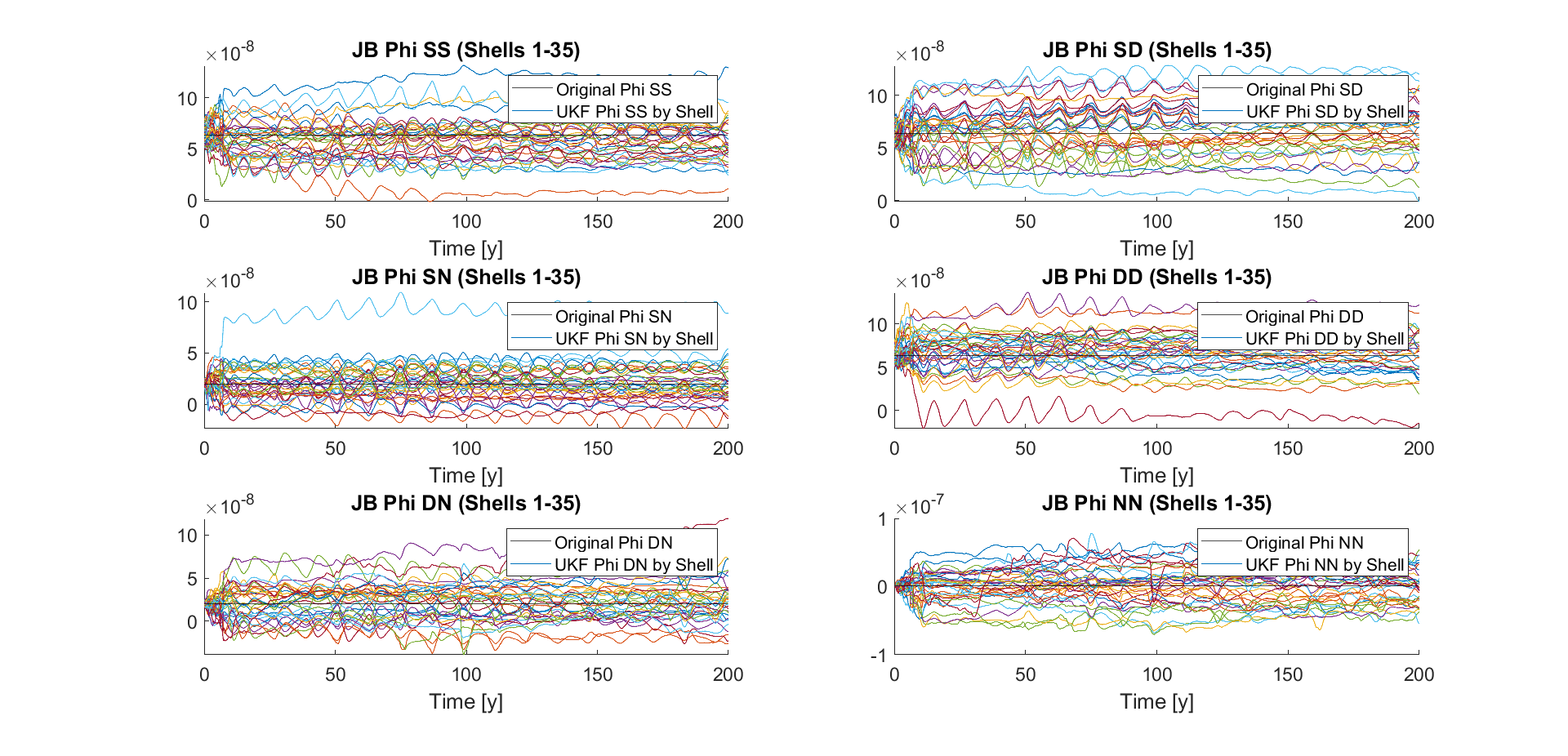}
    \caption{Phi Values Evolution Over Time for Each Shell Using JB2008 Atmospheric Model Without Outliers}
    \label{fig:JBno36}
\end{figure}

\begin{figure}[!ht]
    \centering
    \includegraphics[width=0.99\linewidth]{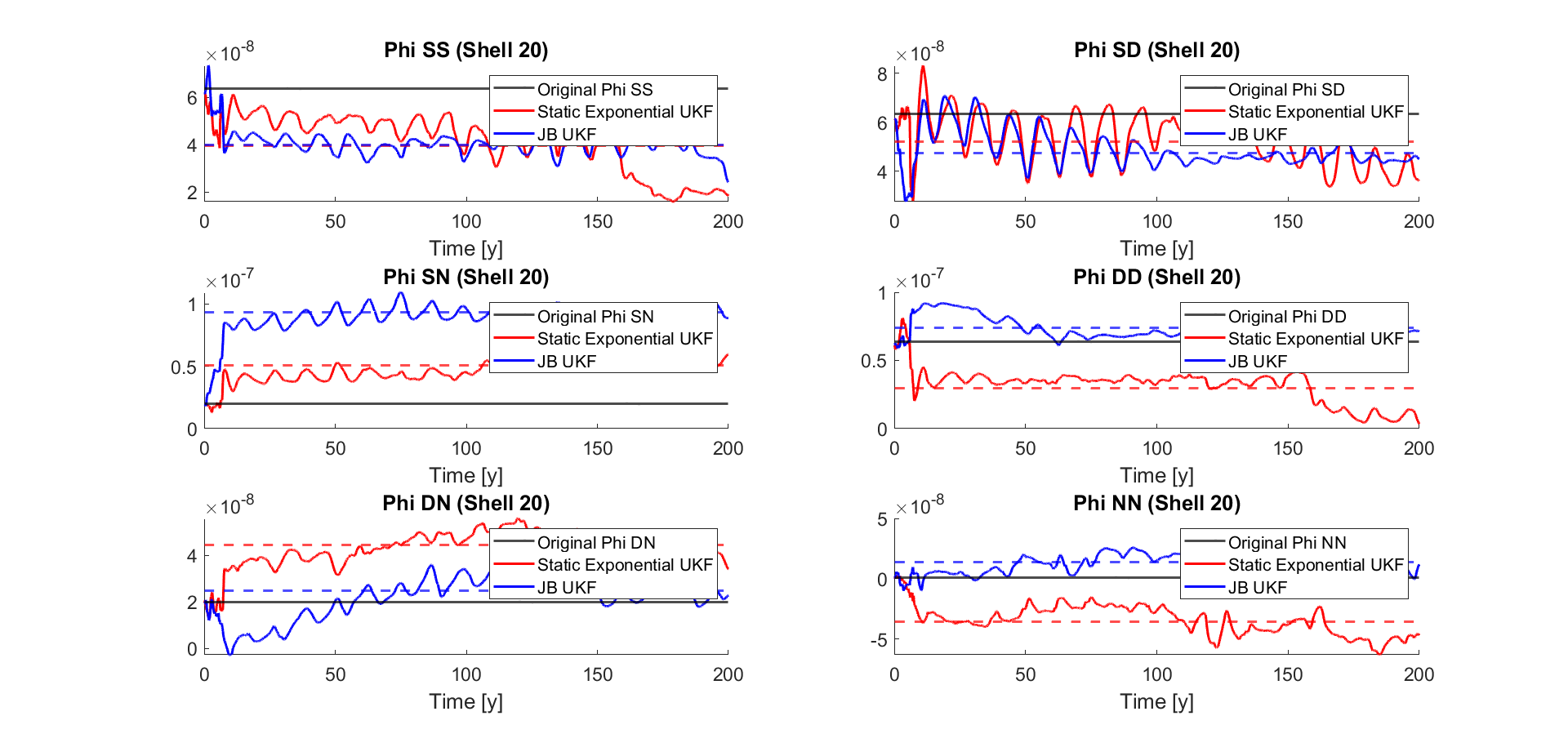}
    \caption{Phi Value Comparison With Steady-State Means (Dashed) In Shell 20 For JB2008 and Static Exponential Atmosphere Models}
    \label{fig:phi20}
\end{figure}

Comparing the evolution of each collision parameter $\phi$, it can be observed that all values connected to S experience an initial significant change from their original values before settling into a steady-state pattern. This is expected behavior, since active control elements on board the spacecraft prevent the collision probability from drastically changing. For the high-fidelity filter, the magnitude of the initial change from original $\phi_{SS}$ values is smaller than those in the low-fidelity filter, albeit with significantly more pronounced oscillation. See Fig. \ref{fig:phi20}.

In contrast, the collision parameters associated with D (excluding $\phi_{SD}$ for the reasons above) consistently show a sinusoidal oscillation over time in both the high-fidelity and low-fidelity filters. The time difference between peaks is approximately twelve years. This shows strong correlation between the collision probability associated with D and the peak solar activity cycle. This is also to be expected as D maintains no active control elements. Objects in D are also significantly more massive and have more surface area than other objects with no active control, thus the effect of the solar cycle is more pronounced.

\begin{table}[]

\caption{$\Phi$ Values*}
\label{tab:phis}
\footnotesize
\centering
\begin{tabular}{ccccccccccccc}
    \toprule \midrule
    \textbf{Shell} & \multicolumn{2}{c}{$\bar{\phi}_{SS}$} & \multicolumn{2}{c}{$\bar{\phi}_{SD}$} & \multicolumn{2}{c}{$\bar{\phi}_{SN}$} & \multicolumn{2}{c}{$\bar{\phi}_{DD}$} & \multicolumn{2}{c}{$\bar{\phi}_{DN}$} & \multicolumn{2}{c}{$\bar{\phi}_{NN}$} \\
    \midrule
     & exp & JB & exp & JB& exp & JB& exp & JB& exp & JB& exp & JB \\
     \midrule

1 & 11.47 & 11.98 & 17.94 & 9.29 & 4.51 & 1.63 & 4.68 & 4.39 & 2.48 & 1.11 & -5.20 & -2.03 \\
2 & 15.33 & 0.76 & 8.87 & 5.96 & 4.43 & 3.55 & 3.94 & 11.53 & -3.22 & 3.97 & 3.12 & -0.39 \\
3 & 6.10 & 3.60 & 12.79 & 5.62 & 6.12 & 2.42 & 4.52 & 8.44 & 3.99 & 3.26 & 4.58 & -0.33 \\
4 & 16.26 & 5.69 & 7.19 & 10.59 & 1.66 & 0.22 & 6.93 & 11.98 & -1.04 & 4.82 & -2.13 & 1.03 \\
5 & 12.91 & 6.99 & 5.94 & 2.56 & 2.93 & 1.55 & 8.87 & 8.36 & 9.63 & 5.80 & 5.91 & -5.09 \\
6 & 12.07 & 7.08 & 3.07 & 12.23 & 6.25 & 4.28 & 7.17 & 5.58 & 0.34 & 1.15 & -4.86 & -4.27 \\
7 & 6.66 & 6.28 & 7.72 & 8.78& 4.89 & 1.55 & 2.57 & 8.99 & -0.80 & 2.22 & 1.71 & 3.04 \\
8 & 12.34 & 8.87 & 10.41 & 2.86 & 1.81 & 3.94 & 4.95 & 5.76 & 0.79 & 1.06 & 0.77 & 4.26 \\
9 & 6.53 & 7.13 & 2.20 & 6.02 & 3.74 & 3.40 & 10.77 & 8.05 & 1.16 & 3.01 & -2.23 & -1.17 \\
10 & 11.77 & 9.06 & 6.85 & 10.19 & 1.29 & 3.46 & 8.88 & 9.33 & -1.49 & 3.15 & 0.74 & -1.26 \\
11 & 9.93 & 6.14 & 7.95 & 7.54 & 1.76 & 2.65 & 7.39 & 6.73 & 1.80 & 0.44 & 4.39 & 0.76 \\
12 & 8.96 & 5.18 & 0.89 & 4.01 & -0.29 & 2.52 & -0.15 & 3.42 & 7.42 & 2.87 & -0.62 & 3.20 \\
13 & 7.49 & 9.98 & 9.35 & 7.51 & -0.32 & 2.15 & 6.44 & 8.65 & 3.27 & 0.27 & -3.13 & -2.35 \\
14 & 4.54 & 7.30 & 1.56 & 10.58 & 2.10 & 2.01 & 1.21 & 9.04 & -2.65 & 3.47 & 2.84 & -0.89 \\
15 & 6.48 & 6.06 & 4.59 & 7.03 & -3.70 & 3.87 & 5.13 & 6.69 & -1.68 & 2.73 & -2.16 & 4.05 \\
16 & 8.35 & 4.26 & 1.48 & 6.37 & 4.01 & 0.59 & 7.80 & 3.00 & -0.45 & -1.71 & -3.06 & -1.01 \\
17 & -0.94 & 7.09 & 8.19 & 5.87 & 1.70 & 1.23 & 9.13 & 6.66 & 3.19 & 3.13 & -2.74 & -4.83 \\
18 & 7.12 & 3.83 & 6.69 & 4.47 & 1.74 & 2.14 & 6.59 & 4.70 & -1.41 & 0.17 & -1.14 & -3.29 \\
19 & 4.56 & 7.22 & 3.99 & 7.07 & 6.09 & 3.42 & 11.04 & 7.73 & 3.22 & 1.75 & -0.39 & -2.89 \\
20 & 3.95 & 3.97 & 5.21 & 4.72 & 5.05 & 9.30 & 2.95 & 7.41 & 4.44 & 2.47 & -3.56 & 1.39 \\
21 & 2.85 & 5.71 & 8.82 & 9.36 & -3.15 & 0.76 & 1.01 & 7.28 & -1.52 & 0.07 & -3.67 & 4.25 \\
22 & 5.42 & 3.34 & 9.07 & 7.41 & -3.74 & 2.03 & 4.84 & 8.20 & 1.15 & 3.97 & 1.81 & 2.08 \\
23 & 4.79 & 5.28 & 6.00 & 5.33 & -0.47 & -1.24 & 7.72 & 6.71 & -1.03 & -1.59 & -0.94 & 2.59 \\
24 & 3.77 & 6.70 & 4.66 & 7.67 & 3.39 & 1.33 & 5.46 & 3.26 & 5.70 & 4.45 & 1.27 & 0.75 \\
25 & 1.40 & 3.84 & 5.21 & 8.62 & 1.13 & 0.77 & 6.15 & 6.85 & -2.78 & 7.61 & 2.50 & 0.86 \\
26 & 5.72 & 3.63 & 3.12 & 5.05 & 3.45 & 3.35 & 5.13 & 5.37 & 1.29 & 1.28 & -2.09 & 0.85 \\
27 & 3.49 & 4.73 & 8.99 & 0.86 & 0.14 & 1.36 & 3.61 & 5.23 & 2.77 & 3.62 & -2.45 & -0.09 \\
28 & 7.53 & 4.26 & 1.19 & 5.48 & 3.87 & 2.88 & 10.39 & 5.63 & -1.84 & 1.39 & 0.25 & 0.07 \\
29 & 6.98 & 6.57 & 5.19 & 8.28 & 0.08 & -0.21 & 7.31 & 5.19 & -3.61 & 4.49 & -3.42 & 1.57 \\
30 & 6.88 & 6.73 & 7.81 & 8.13 & 0.01 & 0.66 & 3.15 & 6.19 & 3.37 & 2.49 & -0.64 & -2.27 \\
31 & 7.50 & 5.62 & 4.57 & 3.88 & 2.90 & 1.31 & 6.44 & 6.40 & -1.77 & 0.66 & -0.78 & 1.39 \\
32 & 6.93 & 7.01 & 7.71 & 2.93 & 3.55 & 0.64 & 7.95 & 7.68 & 0.44 & 0.44 & 0.19 & 1.54 \\
33 & 2.71 & 6.06 & 6.02 & 5.15 & 0.12 & 1.82 & 4.05 & 9.00 & -0.50 & -0.25 & 3.64 & -3.54 \\
34 & 10.32 & 3.17 & 1.77& 1.13 & 5.60 & 4.25 & 7.82 & 6.62 & -4.46 & -0.38 & -1.80 & 4.05 \\
35 & 6.21 & 4.69 & 4.77 & 8.17 & 0.61 & -0.83 & 2.23 & -0.56 & 4.99 & 7.46 & 18.26 & 3.27 \\
36 & 57.88 & 32.67 & 21.36 & 15.17 & -25.26 & -23.81 & 0.38 & -1.87 & 27.78 & 22.33 & 3.41 & 4.55 \\ 
    \midrule 
    \bottomrule 
    \multicolumn{13}{l}{\small * $\Phi$ Values in units of $10^{-8}$.}
\end{tabular}
\end{table}

The sinusoidal oscillation behavior observed in $\phi$ values associated with S and D over time strongly supports the notion that collision probability is significantly affected by peak solar activity cycle.

Since N contains objects from many sources, such as derelict spacecraft collisions, micrometeorites, and discarded launch components, the effect of solar activity on the collision probability of N is not as pronounced due to the wider variation in masses, surface geometries, and sources of each object in N. As a result, the time variation of $\phi_{NN}$ is the most stochastic interaction of all the collision parameters, and most closely resembles a random signal.

Further examination of the evolution of $\phi$ values focused on individual shells. It can be observed in Fig. \ref{fig:JBno36}, that for the majority of shells, the associated $\phi$ values appear to center around the initial value after reaching steady state. However, significant outliers appear for $\phi$ values associated with the uppermost shell as seen in Fig. \ref{fig:JB_phi_all}. We conclude that future modifications to the six augmented $\dot{\phi}$ equations will need to account for difference in shell as well as difference in time relating to solar activity. Steady-state mean $\phi$ values for each shell are given in Table \ref{tab:phis}.

\section{Conclusions}
By incorporating an augmented state vector consisting of each of the three object types in the current MOCAT simulation with six additional collision probability parameters, alongside introducing an Unscented Kalman Filter using Monte-Carlo based measurements, accuracy of the estimate for orbital population is improved. This estimate better conforms to the computationally intensive MOCAT-MC simulation and can be performed in significantly less time.
Data gathered from the update of collision probability estimates will enable future improvements in accuracy as development for a system of ODEs governing the change in collision probability over time continues. Further investigation shows that the accuracy of the UKF may be improved by using gamma distribution fitting which better conforms to the MOCAT-MC data set.

The next stage of this research will be the incorporation of an Extended Kalman Filter (EKF) algorithm and comparison of the results with the UKF. The results of the EKF filter will be further used for data smoothing of the UKF filter.
Future work will also include the development of a Gamma distribution UKF algorithm for better performance and accuracy for this system, as well as incorporating feedback from changes in the collision probability values into the system dynamics. Machine learning will eventually be trained on this process to better anticipate the feedback response.

\section*{Acknowledgments}
The authors wish to acknowledge the support of this work by the MIT MOCAT development team.


\bibliographystyle{AAS_publication}
\bibliography{references}

\end{document}